\documentstyle[amssymb,aps,preprint]{revtex}

\begin{document}
\title{Effects of the Van Hove singularities on magnetism and superconductivity in
the $t$-$t^{\prime }$ Hubbard model: a parquet approach}
\author{V. Yu. Irkhin, A. A. Katanin, and M. I. Katsnelson}
\address{Institute of Metal Physics, 620219 Ekaterinburg, Russia}
\maketitle

\begin{abstract}
The $t$-$t^{\prime }$ Hubbard model for the Fermi level near the Van Hove
singularity is considered within the renormalization group and parquet
approaches. The interplay of ferromagnetic, antiferromagnetic, and
superconducting channels is investigated, and the phase diagram of the model
is constructed. In comparison with previous approaches, the account of
ferromagnetic fluctuations suppresses superconducting pairing and, vice
versa, the influence of the Cooper channel decreases the Curie temperature,
so that the Stoner criterion is inapplicable even qualitatively.
\end{abstract}


\section{Introduction}

Magnetic mechanisms of high-temperature superconductivity (HTSC) have become
the subject of intensive investigations during last decades (see, e.g.,
Refs. \cite{Scalapino,BSW,Pines,Chubukov,Jarrell,Zhang,LK,Morr}). It was
argued that the superconducting properties of HTSC materials are intimately
related to their magnetic properties in the normal phase. In particular,
many features of HTSC compounds were explained from the point of view of the
competition of antiferromagnetic and superconducting order parameters \cite
{Zhang,LK,Morr}. A similar physical situation takes place in ruthenates like
Sr$_2$RuO$_4,$ where the interplay of ferromagnetism and $p$-wave
superconductivity is of crucial importance \cite{Rice}. Both copper-oxide
systems and Sr$_2$RuO$_4$ are layered compounds.\ Therefore a general
problem can be formulated as the investigation of the competition between
magnetic and superconducting instabilities in two-dimensional (2D) electron
systems.

On the other hand, the problem should be concretized. There are some
evidences from both electron structure calculations and experimental data
that the Fermi surface (FS) of HTSC compounds at optimal doping or optimal
pressure is close to the Van Hove (VH) singularities of electron spectrum
(see, e.g., Refs. \cite{mark,abrikos,novikov,andersen}); the situation in
the ruthenates is similar \cite{liebsch}. Due to the presence of VH
singularities, the density of states at the Fermi level becomes
logarithmically divergent, which makes the Fermi liquid unstable with
respect to magnetic ordering or superconductivity. One may expect {\it a
priori} that near VH band fillings the physics is determined by VH points
and is not sensitive to the form of the whole FS. This should be correct
provided that FS is not nested, since in the nesting situation \cite
{Keldysh,Halperin} the contribution of flat FS parts is also important \cite
{Yak,Nest}. As it was first discussed by Dzyaloshinskii \cite{Dzy3}, the
situation in 2D fermion system near VH fillings is similar in some respects
to that in one-dimensional (1D) systems \cite{Dzy1,DzyLar} or in 2D systems
under the nesting condition \cite{Dzy2}. It turned out that in the
instability region the ``normal'' state can demonstrate a non-Fermi liquid
behavior \cite{Dzy96}.

The simplest model which gives a possibility to investigate the effects of
VH singularities on magnetic ordering and superconductivity is the $t$-$%
t^{\prime }$ Hubbard model which takes into account nearest-neighbor and
next-nearest-neighbor hopping. It is often discussed also in connection with
HTSC compounds where the value $t^{\prime }/t=-0.15$ was determined for La$%
_2 $CuO$_4$ and the value $t^{\prime }/t=-0.30$ for the Bi2212 system \cite
{t'/t} (we neglect third-neighbors hopping $t^{\prime \prime }$ which does
not lead to any qualitative changes).

The interplay of antiferromagnetism and superconductivity near VH fillings
within the Hubbard model with $t^{\prime }=0$ was first investigated in
Refs. \cite{Dzy3,DzyYak}. It was shown that the leading instability in this
case is antiferromagnetic one, and the transition temperature is close to
its mean-field value. The authors of Refs. \cite{Dzy3,DzyYak} did not
include the contribution of the particle-hole scattering at small momenta,
as well as particle-particle scattering at momenta near ${\bf Q}=(\pi ,\pi )$
by the reason that these contributions are logarithmic rather than
double-logarithmic. As we will argue in the present paper, even for
weak-to-intermediate coupling regime these contributions should be also
taken into account, which leads to an essential change of the results.

Recently, the authors of Ref. \cite{FurRice} have performed the
renormalization group (RG) analysis of the states close to FS (not only near
VH singularities) within an approach which is similar to that of Ref. \cite
{Shankar}. At $t^{\prime }/t=-0.30$ they found competing antiferromagnetism
and superconductivity, depending on the band filling. However, the approach
of Ref. \cite{Shankar} also does not enable one to treat particle-hole
scattering at small momenta on the equal footing with other contributions.
As it was shown by the RG analysis in Ref. \cite{Guinea}, the account of
particle-hole scattering leads to occurrence of ferromagnetic phase at large
enough $|t^{\prime }|/t;$ the criterion $t^{\prime }/t<-0.27$ for the
stability of ferromagnetism was obtained. At $-0.27<t^{\prime }/t<0$ it was
also found that either antiferromagnetism or superconductivity takes place.
However, unlike Refs.\cite{Dzy3,DzyYak,FurRice}, the contributions of the
Cooper channel were not taken into account in Ref. \cite{Guinea}. The
backward influence of the Cooper channel on magnetic ordering was
investigated within the $T$-matrix approach \cite{Fleck}. It was found for
the non-degenerate Hubbard model that the Cooper channel strongly suppresses
strongly the tendency to ferromagnetism, so that it is possible only for $%
t^{\prime }/t<-0.35.$ Numerical calculations\cite{HSG} predict much larger
values $(t^{\prime }/t)_c$ for the stability of ferromagnetism at VH
filling: $(t^{\prime }/t)_c\gtrsim -0.47$.

Summarizing all these approaches, one can see that we need to consider on
the equal footing all four types of scattering to obtain the correct phase
diagram, i.e. the particle-particle and particle-hole channels at both small
momenta $q$ and ${\bf q}\approx {\bf Q}$. First step in this direction was
made in Refs. \cite{Led,Fur} within the so-called two-patch approach. The
authors of Refs. \cite{Led,Fur} wrote down approximate equations, which were
very similar to the parquet equations in one dimension \cite{Solyom}, and
obtained reasonable physical results. However, they also neglected
particle-hole scattering at small momenta and particle-particle scattering
at ${\bf q}\approx {\bf Q}$ at final stage.

Besides that, these 1D-like equations do not reproduce correctly all the
peculiarities of the 2D dispersion law even close to VH singularities. From
this point of view, most straightforward is the parquet approach of Refs. 
\cite{Dzy3,Dzy1,DzyLar,Dzy2} (for a review see also Ref. \cite{Janis}). It
was applied to the VH singularity\ problem in Refs. \cite{Dzy3,DzyYak}, but
only the case $t^{\prime }/t=0$ was considered (strictly speaking, this case
requires an account of the whole FS because of nesting \cite{Yak}).

In the present paper we consider different phases of the $t$-$t^{\prime }$
Hubbard model $(t^{\prime }/t<0)$ and construct the phase diagram near VH
fillings within the approach \cite{Dzy1}. Note that this approach is
somewhat different from that used in later papers \cite{Dzy2,Dzy3} and, as
we will argue in this work, is more correct.

The outline of the paper is the following. In Sect.2 we discuss
noninteracting susceptibilities and consider the random-phase approximation
(RPA). In Sect. 3 we consider two-patch equations with all four channels of
scattering included and discuss the results of numerical solution of these
equations. In Sect.4 we consider a full parquet approach to VH problem and
compare the results with those of two-patch approach. In conclusion we
summarize main results of the paper and discuss possible directions of
further investigations.

\section{The model and RPA results}

We consider $t$-$t^{\prime }$ Hubbard model on the square lattice: 
\begin{equation}
H=\sum_{{\bf k}}\varepsilon _{{\bf k}}c_{{\bf k}\sigma }^{\dagger }c_{{\bf k}%
\sigma }+U\sum_in_{i\uparrow }n_{i\downarrow }
\end{equation}
with 
\begin{equation}
\varepsilon _{{\bf k}}=-2t(\cos k_x+\cos k_y)+4t^{\prime }\cos k_x\cos
k_y+4t^{\prime }-\mu  \label{ek}
\end{equation}
where $\mu $ is the chemical potential (we have picked out $4t^{\prime }$
for farther convenience) and we have already absorbed the sign of $t^{\prime
}$ into Eq. (\ref{ek}) (hereafter we assume the transfer integrals $%
t,t^{\prime }$ to be positive, $0\leq t^{\prime }/t<1/2$).

The spectrum (\ref{ek}) contains VH singularities connected with the points $%
A=(\pi ,0),$ $B=(0,\pi ).$ These singularities lie at the Fermi surface for
the filling with $\mu =0$ and arbitrary values of $t^{\prime }$. For $%
t^{\prime }=0$ the FS is nested, but the nesting is removed for $t^{\prime
}/t>0.$

Being expanded near the VH singularity points, the spectrum\ (\ref{ek})
takes the form 
\begin{mathletters}
\begin{eqnarray}
\varepsilon _{{\bf k}}^A &=&-2t(\sin ^2\varphi \overline{k}_x^2-\cos
^2\varphi k_y^2)=-2tk_{+}k_{-}-\mu  \label{eka} \\
\varepsilon _{{\bf k}}^B &=&2t(\cos ^2\varphi k_x^2-\sin ^2\varphi \overline{%
k}_y^2)=2t\widetilde{k}_{+}\widetilde{k}_{-}-\mu  \label{ekb}
\end{eqnarray}
where $\overline{k}_x=\pi -k_x,$ $\overline{k}_y=\pi -k_y,$ 
\end{mathletters}
\begin{eqnarray}
k_{\pm } &=&\sin \varphi \overline{k}_x\pm \cos \varphi k_y  \nonumber \\
\widetilde{k}_{\pm } &=&\cos \varphi k_x\pm \sin \varphi \overline{k}_y
\end{eqnarray}
$\varphi $ is the half of the angle between asymptotes at VH singularity, $%
2\varphi =\cos ^{-1}(2t^{\prime }/t).$

At $U=0$ we have the following results for the susceptibilities at small $q$
and ${\bf q}\approx {\bf Q=(}\pi ,\pi {\bf )}$ (see Fig.1a): 
\begin{mathletters}
\label{Hi0}
\begin{eqnarray}
\chi _{{\bf q}}^A &=&\sum_{{\bf k}}\frac{f(\varepsilon _{{\bf k}%
}^A)-f(\varepsilon _{{\bf k+q}}^A)}{\varepsilon _{{\bf k}}^A-\varepsilon _{%
{\bf k+q}}^A}=\frac{z_0}{4\pi ^2t}(\xi _{+}+\xi _{-}),  \label{Hi0a} \\
\chi _{{\bf q+Q}}^{AB} &=&\sum_{{\bf k}}\frac{f(\varepsilon _{{\bf k}%
}^A)-f(\varepsilon _{{\bf k+q}}^B)}{\varepsilon _{{\bf k}}^A-\varepsilon _{%
{\bf k+q}}^B}=\frac 1{2\pi ^2t}\min (z_{{\bf Q}}\xi _{+},z_{{\bf Q}}\xi
_{-},\xi _{+}\xi _{-}).  \label{Hi0b}
\end{eqnarray}
Here $f(\varepsilon )$ is the Fermi distribution function, $\xi _{\pm }=\min
[\ln (\Lambda /q_{\pm }),\ln (\Lambda /\mu )]$%
\end{mathletters}
\begin{equation}
z_0=1/\sqrt{1-R^2};\;z_{{\bf Q}}=\ln [(1+\sqrt{1-R^2})/R],
\end{equation}
and $R=2t^{\prime }/t.$ The expressions for $B\leftrightarrow A$ are obtain
by replacing $\xi _{+}\rightarrow \widetilde{\xi }_{+},\;\xi _{-}\rightarrow
-\widetilde{\xi }_{-}$ where $\widetilde{\xi }_{\pm }=\min [\ln (\Lambda /%
\widetilde{q}_{\pm }),\ln (\Lambda /\mu )].$ The momentum dependence of $%
\chi _{{\bf q}}$ calculated with the spectra (\ref{ek})\ is shown in Fig.2.
Since both the susceptibilities are divergent, we have at least two
competing order parameters. In fact, two other polarization bubbles of
Fig.1b, which are responsible for zero-momentum and $\pi $-pairing, are also
divergent at small ${\bf q}$: 
\begin{mathletters}
\label{Pi0}
\begin{eqnarray}
\Pi _{{\bf q}}^A &=&\sum_{{\bf k}}\frac{1-f(\varepsilon _{{\bf k}%
}^A)-f(\varepsilon _{{\bf k+q}}^A)}{\varepsilon _{{\bf k}}^A+\varepsilon _{%
{\bf k+q}}^A}=\frac{c_0}{2\pi ^2t}\xi _{+}\xi _{-}  \label{Pi0a} \\
\Pi _{{\bf q+Q}}^{AB} &=&\sum_{{\bf k}}\frac{1-f(\varepsilon _{{\bf k}%
}^A)-f(\varepsilon _{{\bf k+q}}^B)}{\varepsilon _{{\bf k}}^A+\varepsilon _{%
{\bf k+q}}^B}=\frac{c_{{\bf Q}}}{2\pi ^2t}\min (\xi _{+},\xi _{-})
\label{Pi0b}
\end{eqnarray}
where 
\end{mathletters}
\[
c_0=1/\sqrt{1-R^2};\;c_{{\bf Q}}=\tan ^{-1}(R/\sqrt{1-R^2})/R 
\]
For $\Pi _{{\bf q}}^B$ we again have the replacements $A\rightarrow B$ and $%
\xi _{+}\rightarrow \widetilde{\xi }_{+},$ $\xi _{-}\rightarrow -\widetilde{%
\xi }_{-}$ in (\ref{Pi0a}).

In the RPA the expressions for particle-hole and particle-particle
susceptibilities read 
\begin{eqnarray}
\overline{\chi }_{{\bf q}} &=&\frac{\chi _{{\bf q}}}{1-U\chi _{{\bf q}}} \\
\overline{\Pi }_{{\bf q}} &=&\frac{\Pi _{{\bf q}}}{1+U\Pi _{{\bf q}}}
\end{eqnarray}
Thus $\Pi $ decreases when the Coulomb interaction is taken into account,
while $\chi $ increases and can diverge at some $U$. In particular, we have
the conventional Stoner criterium of ferromagnetism $U\chi _0=1,$ or 
\begin{equation}
\frac{Uz_0}{2\pi ^2t}\ln \frac \Lambda \rho =1
\end{equation}
where $\rho =\max (T/t,\mu /t,\Delta /t)$ ($\Delta \sim \overline{S}$ is the
spin splitting). The solution to this equation reads 
\[
\rho =\Lambda \exp \left[ -2\pi ^2\sqrt{t^2-(2t^{\prime })^2}/U\right] 
\]
Therefore the ferromagnetism is present at any $U;$ moreover, at $U\sim 2\pi
^2[t^2-(2t^{\prime })^2]^{1/2}$ one can expect that it becomes saturated.
Similarly, considering the antiferromagnetic instability we obtain 
\begin{equation}
\frac U{2\pi ^2t}\min (\ln ^2\frac \Lambda \rho ,z_{{\bf Q}}\ln \frac \Lambda
\rho )=1
\end{equation}
which gives 
\[
\rho =\Lambda \left\{ 
\begin{array}{cc}
\exp (-\sqrt{2\pi ^2t/U}), & U/(2\pi ^2t)>1/z_{{\bf Q}}^2 \\ 
\exp (-2\pi ^2t/z_{{\bf Q}}U), & U/(2\pi ^2t)<1/z_{{\bf Q}}^2
\end{array}
\right. 
\]
so that antiferromagnetism is favorable at small $t^{\prime }/t.$

However, as it was discussed first by Dzyaloshinskii and coworkers \cite
{Dzy3,Dzy1,DzyLar,Dzy2}, RPA is incorrect even in the weak-coupling limit,
except for the case when only one bubble is divergent. Since in the VH case
all four bubbles of Fig.1 are divergent, we have to use the parquet approach 
\cite{Dzy3,Dzy1,DzyLar,Dzy2} instead of RPA. While in 1D case the parquet
equations reduce to conventional differential RG equations (see, e.g., Refs. 
\cite{Yak,DzyLar,Solyom}), for higher space dimensionalities we have coupled
integral equations. First we consider the approach of Refs. \cite{Led,Fur}
which uses mapping of the full parquet equations on an ``effective'' 1D
problem, i.e. so-called two-patch equations.

\section{Two-patch equations}

The authors of Refs.\cite{Led,Fur} proposed the approach which neglects the
difference between $\xi _{+}$ and $\xi _{-}$ (and consequently between $%
\widetilde{\xi }_{+}$ and $\widetilde{\xi }_{-}$) and introduced a single
scaling variable $\xi =\min (\xi _{+},\xi _{-},\widetilde{\xi }_{+},%
\widetilde{\xi }_{-}).$ Note that this approach is not strict, in particular
because of the presence of double-logarithmic terms in (\ref{Hi0b}) and (\ref
{Pi0a}). At the same time, as we will see below (see also Ref. \cite{Fur}),
this reproduces correctly main features of the exact parquet equations.

The two-patch equations read \cite{Led,Fur} 
\begin{eqnarray}
\gamma _1^{\prime } &=&2d_1(\xi )\gamma _1(\gamma _2-\gamma _1)+2d_2\gamma
_1\gamma _4-2\,d_3\gamma _1\gamma _2  \nonumber \\
\gamma _2^{\prime } &=&d_1(\xi )(\gamma _2^2+\gamma _3^2)+2d_2(\gamma
_1-\gamma _2)\gamma _4-d_3(\gamma _1^2+\gamma _2^2)  \nonumber \\
\gamma _3^{\prime } &=&-2d_0(\xi )\gamma _3\gamma _4+2d_1(\xi )\gamma
_3(2\gamma _2-\gamma _1)  \nonumber \\
\gamma _4^{\prime } &=&-d_0(\xi )(\gamma _3^2+\gamma _4^2)+d_2(\gamma
_1^2+2\gamma _1\gamma _2-2\gamma _2^2+\gamma _4^2)  \label{TwoPatch}
\end{eqnarray}
where $\gamma _i^{\prime }\equiv d\gamma _i/d\xi $, 
\begin{eqnarray}
d_0(\xi ) &=&2c_0\xi ;\;d_1(\xi )=2\min (\xi ,z_Q)  \nonumber \\
d_2 &=&2z_0;\;d_3=2c_{{\bf Q}}
\end{eqnarray}
and four vertices $\gamma _{1-4}$ are defined in Fig.3. In these notations, $%
\gamma _i(0)=g_0\equiv U/(4\pi ^2t)$ corresponds to the Hubbard model. While
only the case $d_2,d_3\ll d_0,d_1$ was considered in Refs. \cite{Led,Fur},
we perform a more general consideration where all the bubbles are taken into
account. We have also taken into account the coefficient $c_0$ to treat
correctly the $t^{\prime }$ dependence of the amplitude of particle-particle
scattering. Note that the equations (\ref{TwoPatch}) are very similar to
those in the 1D case \cite{Yak,DzyLar,Solyom} with the difference that in
the latter case one has 
\begin{equation}
d_0=d_2=0;\;d_1=d_3=1
\end{equation}
The complete discussion of the physics of the equations (\ref{TwoPatch}) in
the 1D case is given in Ref. \cite{DzyLar}. In two dimensions, the
coefficients $d_0$ and $d_1$ become $\xi $-dependent because of the presence
of double-logarithmic terms. As we have already mentioned, this gives only
approximate treatment of such terms. The equations (\ref{TwoPatch}) give a
possibility to investigate the interplay of all the four scattering channels.

For ferromagnetic, antiferromagnetic, and d-wave superconducting
susceptibilities we have 
\begin{equation}
\chi _{\text{F},\text{AF},d\text{-SC}}(\xi )=\int\limits_0^\xi d\zeta
d_{2,1,0}(\zeta ){\cal T}_{\text{F},\text{AF},d\text{-SC}}^2(\zeta )
\end{equation}
where ${\cal T}$ satisfies the equation 
\begin{equation}
\frac{d\ln {\cal T}_{\text{F},\text{AF},d\text{-SC}}}{d\xi }=\left\{ 
\begin{array}{c}
d_2(\gamma _1+\gamma _4) \\ 
d_1(\xi )(\gamma _2+\gamma _3) \\ 
d_0(\xi )(\gamma _3-\gamma _4)
\end{array}
\right\}  \label{T}
\end{equation}
Thus, when $\gamma _1$ and $\gamma _4$ are simultaneously relevant, we have
ferromagnetic ordering, while $\gamma _2$ and $\gamma _3$ lead to
antiferromagnetic ordering. For the superconductivity, we have a more
complicated combination of relevant and irrelevant vertices.

The results of the solution of equations (\ref{TwoPatch})-(\ref{T}) for
various values of $g_0$ and $t^{\prime }/t$ are shown in Fig.4. Depending on
the values of $g_0$ and $t^{\prime }/t,$ ferro- or antiferromagnetic
susceptibility, or $d$-wave superconducting response diverges first. The
parameter dependences of the critical energy scale $\mu _c=t\exp (-\xi _c)$
are shown in Fig.5. This scale can be approximately identified with the
critical chemical potential or transition temperature. For comparison, the
RPA results for the stability of ferro- and antiferromagnetism are shown
too. One can see that the values of transition temperatures obtained from
the two-patch equations are much lower than the corresponding RPA results.

To understand qualitatively the nature of the critical temperature lowering,
we may neglect the interpatch scattering. In the ferromagnetic case we have
only one nonzero vertex $\gamma _4,$ and the equation for it has the form 
\begin{equation}
\gamma _4^{\prime }=-2(z_0-c_0\xi )\gamma _4^2
\end{equation}
so that 
\begin{equation}
\gamma _4=\frac{g_0}{1+g_0(c_0\xi ^2-2z_0\xi )}  \label{g4}
\end{equation}
The modification of the Stoner criterion takes the form 
\begin{equation}
g_0(2z_0\ln \frac \Lambda \mu -c_0\ln ^2\frac \Lambda \mu )=1  \label{MSC}
\end{equation}
Then we have from (\ref{MSC}) at $z_0=c_0$ (which is the case of $t$-$%
t^{\prime }$ model) 
\[
\ln \frac \Lambda \mu =1-\frac{\sqrt{z_0g_0-1}}{z_0g_0}>\frac 1{2z_0g_0} 
\]
(for $R$ close to unity we have $z_0g_0>1$). Thus, the decrease of the Curie
temperature in comparison with the mean-field\ approach ($c_0=0$) is
directly connected with the account of the Cooper bubble, which is in
agreement with the $T$-matrix approximation\cite{Fleck}. Note, however, that
the structure of Eq. (\ref{g4}) is different from that obtained in the $T$%
-matrix approach.

The resulting phase diagram in $U$-$t^{\prime }/t$ plane with all the
scattering channels being included is shown in Fig. 6. One can see that the $%
d$-wave superconducting response is strongly suppressed by particle-hole
scattering processes. This is the leading divergent response only at small
values of coupling constant $g_0<0.04,$ which corresponds to $U<1.6t$. The
critical temperature in this region is also exponentially small ($T_c\sim
\mu _c\sim \exp (-1/g_0)$).

\section{Parquet equations}

Now we pass to the consideration of the full parquet equations and compare
the results of their solution with approximate RG equations of Sect. 2. We
use the generalization of the approach of Ref. \cite{Dzy1} to the case of
two dimensions.

In the parquet approach (see Appendix) we have for each vertex $i=1...4$
three types of bricks, which are shown in Fig.7: the Cooper brick $C_i(\xi
_{\pm },\eta _{\pm }),$ and two zero-sound bricks, $Z_i(\xi _{\pm },\eta
_{\pm })$ and $\widetilde{Z}_i(\xi _{\pm },\eta _{\pm })$. Up to a
logarithmic accuracy, they depend on $\xi _{\pm }=\ln (\Lambda /k_{\pm })$
and $\eta _{\pm }=\ln (\Lambda /q_{\pm })$ only, $k_{\pm }=k_{1\pm }+k_{2\pm
}$ and $q_{\pm }=\max \{k_{3\pm }-k_{1\pm },k_{3\pm }-k_{2\pm }\}$ being the
Cooper and zero-sound momenta transfer. The vertices $\gamma _i(\xi _{\pm
},\eta _{\pm })$ in different regions of $\xi _{\pm }$ and $\eta _{\pm }$
are given by \cite{Dzy1} 
\begin{eqnarray}
\gamma _i(\xi _{\pm },\eta _{\pm }) &=&\gamma _i^h(\xi _{\pm },\eta _{\pm
})\equiv g_0+C_i(\xi _{\pm },\eta _{\pm })+Z_i(\eta _{\pm },\eta _{\pm })+%
\widetilde{Z}_i(\eta _{\pm },\eta _{\pm })\;(\xi _{\pm }>\eta _{\pm }) 
\nonumber \\
\gamma _i(\xi _{\pm },\eta _{\pm }) &=&\gamma _i^l(\xi _{\pm },\eta _{\pm
})\equiv g_0+C_i(\xi _{\pm },\xi _{\pm })+Z_i(\xi _{\pm },\eta _{\pm })+%
\widetilde{Z}_i(\xi _{\pm },\xi _{\pm })\;(\xi _{\pm }<\eta _{\pm },\eta
_{\pm }^{(1)}<\eta _{\pm }^{(2)})  \nonumber \\
\gamma _i(\xi _{\pm },\eta _{\pm }) &=&\widetilde{\gamma }_i^l(\xi _{\pm
},\eta _{\pm })\equiv g_0+C_i(\xi _{\pm },\xi _{\pm })+Z_i(\xi _{\pm },\xi
_{\pm })+\widetilde{Z}_i(\xi _{\pm },\eta _{\pm })\;(\xi _{\pm }<\eta _{\pm
},\eta _{\pm }^{(1)}>\eta _{\pm }^{(2)})  \label{G}
\end{eqnarray}
where $\eta _{\pm }^{(1,2)}=\ln (\Lambda /|k_{3\pm }-k_{1,2\pm }|).$
Following to Ref. \cite{Dzy1}, we have taken into account that at $\xi _{\pm
}>\eta _{\pm }$ the Cooper brick depends on both $\xi _{\pm }$\ and $\eta
_{\pm },$ while the zero-sound bricks depend only on $\eta _{\pm }.$ Vice
versa, at $\xi _{\pm }<\eta _{\pm }$ the Cooper brick and one of the
zero-sound bricks depend only on $\xi _{\pm },$ and another zero-sound brick
depends on both $\xi _{\pm }$\ and $\eta _{\pm }.$

When all momenta are of the same order, i.e. $\xi _{\pm }=\eta _{\pm },$ the
vertices 
\begin{equation}
\gamma _i(\xi _{\pm },\xi _{\pm })=\gamma _i(\xi _{\pm })
\end{equation}
are analogous to those introduced in Sect.3 with the only difference that
now they depend on two scaling variables $\xi _{\pm }$. However, unlike the
1D case, the parquet equations do not reduce to the equations for $\gamma
_i(\xi _{\pm }),$ but contain the full dependence $\gamma _i(\xi _{\pm
},\eta _{\pm }).$ The corresponding equations are presented in Appendix. As
discussed in Appendix, the approach we use gives a possibility to treat the
2D situation in a more correct way in comparison with the approach of Refs. 
\cite{Dzy2,Dzy3}.

The parquet equations were solved numerically. To this end, we placed the
variables $\xi ,\eta $ on a grid with $16$ points in each dimension, so that
the total number of vertices to be taken into account is $3\cdot 4\cdot
16^4\approx 8\cdot 10^6$. It is important that the grid was chosen for the
logarithmic variables $\xi ,\eta ,$ but not for the momenta themselves. This
gives a possibility to use simple integration methods (e.g., the trapezium
method) to obtain the results which are correct to logarithmic accuracy. The
resulting system of $8\cdot 10^6$ algebraic equations was solved by the
Zeidel method.

The structure of the solutions of the parquet equations is quite similar to
that in the two-patch approach, except for that now we have
momenta-dependent vertices. Again, the relevance of $\gamma _1$ and $\gamma
_4$ with $\xi _{\pm }=\eta _{\pm }=\xi $ leads to ferromagnetic ordering,
while the relevance of $\gamma _2$ and $\gamma _3$ at $\xi _{\pm }=\eta
_{\pm }=\xi $ to antiferromagnetic one. The results of solution of the
parquet equations are shown in Figs.8, 9. One can see that the results
coincide qualitatively with those of the two-patch parquet approach of
Sect.3. At not too large $t^{\prime }/t,$ the antiferromagnetic instability
occurs first, while for $t^{\prime }/t$ close to $1/2$ the leading
instability is ferromagnetic one. The superconductivity occurs also only for
very small $g_0$.

The transition temperatures obtained within the parquet approach are larger
than those obtained from two-patch equations, but are still lower than the
RPA results. In particular, in the limit of small $t^{\prime }/t$ the
parquet calculations\cite{Dzy3,DzyYak}, which do not take into account
single-logarithmic contributions of the loops Fig.1a,d, yield for $g_0=0.1$
the critical value for stability of antiferromagnetism $\xi _c^2=5.2,$ which
is close to RPA result, $\xi _c^2=5.0$. At the same time, our parquet
calculations give larger value, $\xi _c^2=6.37$ (the result of two-patch
equations is $\xi _c^2=18.2$). The region of stability of $d$-wave
superconducting phase is even smaller than that obtained from two-patch
equations.

The critical concentrations $n_c$ for the stability of ferro- or
antiferromagnetic and superconducting phases close to VH filling can be
estimated from the critical chemical potential with the use of the condition 
\begin{equation}
n=\sum_{{\bf k}}f(\varepsilon _{{\bf k}})
\end{equation}
Using the form of spectrum (\ref{ek}) and taking the limit of filling close
to VH one, we obtain for $|\mu |\ll t$ 
\begin{equation}
\delta n_c=n_c-n_{\text{VH}}=\frac{\mu _c}{2\pi ^2t\sqrt{1-R^2}}\ln \frac{%
\Lambda t}{|\mu _c|}\simeq \frac{\xi _c\exp (-\xi _c)}{2\pi ^2\sqrt{1-R^2}}
\end{equation}
where $n_{\text{VH}}$ is the VH filling. In particular, for $g_0=0.1$ ($%
U=3.95t$) we have from Fig.8 
\begin{eqnarray}
\delta n_c &=&0.01\;(\text{AF phase, }t^{\prime }/t\rightarrow 0)  \nonumber
\\
\delta n_c &=&0.03\;(\text{F phase,\ }t^{\prime }/t=0.45)
\end{eqnarray}
Thus, except for the limit $R\rightarrow 1$ ($t^{\prime }\rightarrow t/2$),
the critical concentrations are very small, which is in qualitative
agreement with the results of Ref. \cite{Fleck}. Because of the exponential
smallness of the critical chemical potential, the critical concentrations
for the superconducting phase are even smaller than those for the
magnetically ordered phases.

\section{Conclusion}

Now we summarize the main results of the paper. Using the two-patch
equations (Sect. 3) and parquet equations (Sect. 4) we constructed the phase
diagrams of $t$-$t^{\prime }$ Hubbard model (Figs. 5,6,8,9) at the fillings
which are close to Van Hove one. It was argued that the simultaneous account
of all the scattering channels is important in considering the VH problem,
the smallness of contributions of some channels (logarithmical vs.
double-logarithmical divergence) being compensated by the growth of relevant
couplings. Both the approaches used, two-patch and parquet ones, give
similar phase diagrams. In agreement with the previous approaches \cite
{Guinea,Fleck,HSG}, antiferromagnetism is favorable for small $t^{\prime
}/t, $ while ferromagnetism for larger values of $t^{\prime }/t.$ The
stability of antiferro- and especially ferromagnetism is greatly reduced in
comparison with the corresponding mean-field criteria. Thus the Stoner
criterion is completely inapplicable for the systems with VH singularities;
depending on the value $t^{\prime }/t,$ it overestimates the critical
temperature by 2-10 times. This conclusion is in qualitative agreement with
the results of Ref. \cite{Fleck}. Besides that, the mean-field approach is
unable to determine the critical value $(t^{\prime }/t)_c$ which separates
the ferro- and antiferromagnetic phases.

Unlike Ref. \cite{Guinea}, $(t^{\prime }/t)_c$ is $U$-dependent and
decreases with increasing $U.$ Although the RG (and also parquet) approach
is unable to describe the ordered states, from scaling arguments we have $%
\overline{S}\propto (\mu _c/t)^\beta $, where $\beta $ is the magnetization
critical exponent. With increasing $U,$ the ferro- and antiferromagnetic
states are characterized by large magnetic moments, and ferromagnetism
possibly becomes saturated. However, these values of $U$ are not described
by perturbative approaches and should be treated in the strong-coupling
limit. At the same time, determining parameters of the
ferro-antiferromagnetic quantum phase transition would be of interest,
especially the critical exponents. One can expect that they are independent
of the coupling.

Another result of the paper is that the tendency to $d$-wave superconducting
pairing is considerably reduced in comparison with the treatments of Refs. 
\cite{Guinea,FurRice}: it can occur only at very small values of $U$. Of
course, this concerns only the pairing due to the VH singularities
themselves; the pairing can be further enhanced by other factors. This can
be also the subject for future investigations. Details of the electron
spectrum, especially the form of the fermion Green's functions close to the
phase transition into the ferro- or antiferromagnetic state are beyond the
scope of the present paper. Although the marginal \cite{MFL} and
non-Fermi-liquid behavior \cite{Dzy96} was found (see also the discussion in
Ref. \cite{FLP}), this problem needs further investigations, since a
simultaneous account of all the scattering channels can be important in this
case too. For example, only one of four scattering channels was included in
Ref. \cite{Dzy96}.

We believe that the results of the present paper can be also important for
the theory of itinerant-electron ferromagnetism. A standard consideration
(including contemporary spin-fluctuation theories \cite{Moriya}) starts from
the RPA approach. It was noted in Ref. \cite{IKT} that for almost all known
itinerant electron ferromagnets the Fermi level lies near a 2D-like VH
singularity. This is a result of merging two weaker 3D square-root
singularities along symmetrical directions in the Brillouin zone \cite{Kats}%
. We have shown that under such conditions the RPA approach and the Stoner
criterion are not applicable even qualitatively because of the strong
interference with the Cooper channel. Of course, the effect of the
logarithmic VH singularity in the 3D case is not exactly the same as in the
pure 2D case considered here, so that the 3D problem needs further
investigations. However, the naive Stoner criterion is in any case doubtful
and needs a careful justification.

In this respect, it would be interesting to generalize the results of the
present paper (at least those from the two-patch equations) on the
degenerate-band Hubbard model. As was argued in Ref. \cite{Fleck}, in this
case the suppression of ferromagnetic ordering is much weaker than for the
nondegenerate model considered. One can also expect that the particle-hole
scattering with small momenta will not renormalize superconducting channel
as strongly as for the non-degenerate model. However, these statements need
further justification since the diagram series in the degenerate and
non-degenerate cases look like rather similar.

The research described was supported in part by Grant No.00-15-96544 from
the Russian Basic Research Foundation (Support of Scientific Schools).

\appendix

\section{The parquet approach in one and two dimensions}

First we consider the simple model of spinless fermions 
\begin{equation}
H=\sum_{{\bf k}}\varepsilon _{{\bf k}}c_{{\bf k}}^{\dagger }c_{{\bf k}%
}+\sum_{{\bf k,p}}g({\bf p})c_{{\bf k}}^{\dagger }c_{{\bf k-p}}c_{{\bf k}%
}^{\dagger }c_{{\bf k+p}}
\end{equation}
with $g(k_F)=g_0.$ In one dimension we have the representation \cite{Dzy1}
for the renormalized vertex $\gamma $ (Fig. 5) 
\begin{equation}
\gamma (\xi ,\eta )=g_0+C(\xi ,\eta )+Z_1(\xi ,\eta )+Z_2(\xi ,\eta )
\label{g1}
\end{equation}
where the bricks are given by 
\begin{eqnarray}
C(\xi ,\eta ) &=&-c\int\limits_0^\xi d\zeta \gamma ^c(\overline{\zeta ,\eta }%
)\gamma ^h(\zeta ,\eta ),\;\xi >\eta  \nonumber \\
Z_{1,2}(\xi ,\eta ) &=&z_{1,2}g_0\int\limits_0^\eta d\zeta \gamma _{1,2}^z(%
\overline{\zeta ,\xi })\gamma _{1,2}^l(\zeta ,\eta ),\;\xi <\eta  \label{CZ1}
\end{eqnarray}
and $\xi =\ln (\Lambda /|k_1+k_2|)$; $\eta =\ln (\Lambda /\max
\{k_3-k_1,k_3-k_2\}).$ Here 
\begin{equation}
\overline{\zeta ,\xi }=\left\{ 
\begin{array}{cc}
\zeta & |\zeta |<|\xi | \\ 
\xi & |\zeta |>|\xi |
\end{array}
\right. ,
\end{equation}
and we assume that the Cooper, ZS and ZS$^{\prime }$ loops are
logarithmically divergent with the coefficients $c$ and $z_{1,2}$
respectively (we generalize here the approach of Ref. \cite{Dzy1} to the
case where both the channels, ZS and ZS$^{\prime }$ contain divergences).
The vertices in (\ref{CZ1}) are given by 
\begin{eqnarray}
\gamma ^c(\xi ) &=&g_0+Z_1(\xi ,\xi )+Z_2(\xi ,\xi )  \nonumber \\
\gamma _{1,2}^z(\xi ) &=&g_0+C(\xi ,\xi )+Z_{2,1}(\xi ,\xi )  \label{gcz1} \\
\gamma ^h(\xi ,\eta ) &=&\gamma ^c(\xi )+C(\xi ,\eta )  \nonumber \\
\gamma _{1,2}^l(\xi ,\eta ) &=&\gamma _{1,2}^z(\xi )+Z_{1,2}(\xi ,\eta ) 
\nonumber
\end{eqnarray}

The equations (\ref{g1})-(\ref{gcz1}) form the closed system of parquet
equations for the 1D spinless case. The validity of these equations can be
demonstrated for the trivial case $z_1=z_2=0$ where the direct ladder (RPA)
summation is possible. In this case $Z_{1,2}(\xi ,\eta )=0,\;\gamma ^c(\xi
)=g_0,$ and we obtain from (\ref{g1}), (\ref{gcz1}) the standard ladder
equation 
\begin{equation}
\gamma (\xi ,\eta )=g_0-cg_0\xi \gamma (\xi ,\eta )
\end{equation}
This has the solution 
\begin{equation}
\gamma (\xi ,\eta )=\frac{g_0}{1+cg_0\xi }
\end{equation}

Now we return to the general case $z_1,z_2,c\neq 0.$ As it is shown in Ref. 
\cite{Dzy1}, the equations (\ref{g1})-(\ref{gcz1}) at $\xi =\eta $ can be
reduced to 
\begin{equation}
\gamma (\xi ,\xi )\equiv \gamma (\xi )=g_0+(z_1+z_2-c_0)\int\limits_0^\xi
\gamma ^2(\zeta )d\zeta  \label{RG}
\end{equation}
which is equivalent to the differential RG equation 
\begin{equation}
\frac{d\gamma }{d\xi }=(z_1+z_2-c)\gamma ^2  \label{RG1}
\end{equation}
Since in one dimension $z_1=c,$ $z_2=0,$ $\gamma $ is marginal and we have a
Luttinger-liquid behavior \cite{Shankar,Solyom,Yak}. Alternatively, the
equations (\ref{RG})\ or (\ref{RG1}) can be obtained directly with the use
of the Sudakov's trick \cite{Sudakov} or standard RG approach \cite{Shankar}
without considering the general dependence $\gamma (\xi ,\eta )$. Thus, in
the 1D case the parquet and RG approaches are equivalent.

In the 2D case we have two pairs of variables, $\xi _{\pm }$ and $\eta _{\pm
}.$ However, unlike the 1D case, now two possibilities occur: the momentum
integration in bubbles can be logarithmical or double-logarithmical. For
example we consider the case where the integration in the Cooper bubble is
double-logarithmic while in the zero-sound channel this yields only simple
logarithms (which is similar to the situation for VH singularities). Then it
can be checked by a direct comparison with perturbation theory that the
equations 
\begin{eqnarray}
C(\xi _{\pm },\eta _{\pm }) &=&-cg_0\int\limits_0^{\xi
_{+}}\int\limits_0^{\xi _{-}}d\zeta _{+}d\zeta _{-}\gamma ^c(\overline{\zeta
_{\pm },\eta _{\pm }})\gamma ^h(\zeta _{\pm };\eta _{\pm }),\;\xi _{\pm
}>\eta _{\pm }  \nonumber \\
Z_{1,2}(\xi _{\pm },\eta _{\pm }) &=&z_{1,2}g_0\int\limits_0^{\eta
_{+}}d\zeta _{+}\gamma _{1,2}^z(\overline{\zeta _{+},\xi _{+}},\xi
_{-})\gamma _{1,2}^l(\zeta _{+},\eta _{-};\eta _{\pm })  \nonumber \\
&&\ \ \ +z_{1,2}g_0\int\limits_0^{\eta _{-}}d\zeta _{-}\gamma _{1,2}^z(\xi
_{+},\overline{\zeta _{-},\xi _{-}})\gamma _{1,2}^l(\eta _{+},\zeta
_{-};\eta _{\pm }),\;\xi _{\pm } 
\begin{array}{c}
<
\end{array}
\eta _{\pm }  \label{CZ2}
\end{eqnarray}
with 
\begin{eqnarray}
\gamma ^c(\eta _{\pm }) &=&g_0+Z_1(\eta _{\pm },\eta _{\pm })+Z_2(\eta _{\pm
},\eta _{\pm })  \nonumber \\
\gamma _i^z(\xi _{\pm }) &=&g_0+C_i(\xi _{\pm },\xi _{\pm })+Z_{3-i}(\xi
_{\pm },\xi _{\pm })  \nonumber \\
\gamma ^h(\xi _{\pm },\eta _{\pm }) &=&\gamma ^c(\eta _{\pm })+C(\xi _{\pm
},\eta _{\pm })  \nonumber \\
\gamma _i^l(\xi _{\pm },\eta _{\pm }) &=&\gamma _i^z(\xi _{\pm })+Z_i(\xi
_{\pm },\eta _{\pm })  \label{G2}
\end{eqnarray}
give the parquet solution of the problem. Note that beyond one dimension the
integral parquet equations do not reduce to differential ones.

The above approach is different from the approach of Refs. \cite
{Dzy2,Dzy3,Yak} where a standard RG scheme was applied in one dimension,
while momentum dependence in another dimension was taken into account
exactly rather than to logarithmic accuracy. However, the applicability of
last approach is doubtful. Indeed, in the 1D case the equation (\ref{RG})
can be considered as a logarithmic approximation to the Bethe-Salpeter
equations 
\begin{eqnarray}
C(k_1,k_2,k_3) &=&g_0-c_0g_0\int\limits_0^\Lambda dk\gamma (k_1+k,k_2-k,k_3)
\nonumber \\
Z_1(k_1,k_2,k_3) &=&g_0+z_1g_0\int\limits_0^\Lambda dk\gamma (k_1,k_3+k,k_3)
\nonumber \\
Z_2(k_1,k_2,k_3) &=&g_0+z_2g_0\int\limits_0^\Lambda dk\gamma
(k_1,k_2+k,k_3+k)  \label{BS}
\end{eqnarray}
If we have {\it both }slow and fast momenta we need to combine (\ref{RG})
and (\ref{BS}) which is impossible since (\ref{RG}) is quadratic in $\gamma $
while (\ref{BS}) is {\it linear}. The equations of Refs. \cite{Dzy2,Dzy3,Yak}
\begin{eqnarray}
C(k_1,k_2,k_3,\xi ) &=&g_0-c_0\int\limits_0^\xi d\zeta \int\limits_0^\Lambda
dk\gamma (k_1,k_2,k_3+k;\zeta )\gamma (k_1+k,k_2-k,k_3;\zeta )  \nonumber \\
Z_1(k_1,k_2,k_3,\xi ) &=&g_0+z_1\int\limits_0^\xi d\zeta
\int\limits_0^\Lambda dk\gamma (k_1,k_3+k,k_3;\zeta )\gamma
(k_1+k,k_2,k_3+k;\zeta )  \nonumber \\
Z_2(k_1,k_2,k_3,\xi ) &=&g_0+z_2\int\limits_0^\xi d\zeta
\int\limits_0^\Lambda dk\gamma (k_1,k_2+k,k_3+k;\zeta )\gamma
(k_3+k,k_2,k_3;\zeta )
\end{eqnarray}
are not fully correct. If we suppose that $\gamma $ does not depend on $\xi $%
, we do not reproduce the 1D Bethe-Salpeter equations (\ref{BS}). At the
same time, the approach of Ref. \cite{Dzy1} is free from these problems.

The generalization of equations (\ref{CZ2}) to the full VH problem is
trivial. The parquet equations have the form

\begin{eqnarray*}
C_1(\xi _{\pm };\eta _{\pm }) &=&-c_{{\bf Q}}\int_{c_{{\bf Q}}}\left[ \gamma
_1^c(\overline{\zeta _{\pm }\eta _{\pm }})\gamma _2^h(\xi _{\pm };\zeta
_{\pm })+\gamma _2^c(\overline{\zeta _{\pm }\eta _{\pm }})\gamma _1^h(\xi
_{\pm };\zeta _{\pm })\right] \\
Z_1(\xi _{\pm };\eta _{\pm }) &=&z_{{\bf Q}}\int_{z_{{\bf Q}}}\left[ \gamma
_1^z(\overline{\zeta _{\pm }\xi _{\pm }})\widetilde{\gamma }_2^l(\xi _{\pm
};\zeta _{\pm })+\widetilde{\gamma }_2^z(\overline{\zeta _{\pm }\xi _{\pm }}%
)\gamma _1^l(\xi _{\pm };\zeta _{\pm })-2\gamma _1^z(\overline{\zeta _{\pm
}\xi _{\pm }})\gamma _1^l(\xi _{\pm };\zeta _{\pm })\right. \\
&&\ \ \ \ \ \left. +\gamma _3^z(\overline{\zeta _{\pm }\xi _{\pm }})%
\widetilde{\gamma }_3^l(\xi _{\pm };\zeta _{\pm })+\widetilde{\gamma }_3^z(%
\overline{\zeta _{\pm }\xi _{\pm }})\gamma _3^l(\xi _{\pm };\zeta _{\pm
})-2\gamma _3^z(\overline{\zeta _{\pm }\xi _{\pm }})\gamma _3^l(\xi _{\pm
};\zeta _{\pm }))\right] \\
\widetilde{Z}_1(\xi _{\pm };\eta _{\pm }) &=&z_0\int_{z_0}^A\widetilde{%
\gamma }_4^z(\overline{\zeta _{\pm }\xi _{\pm }})\widetilde{\gamma }%
_1^l(\zeta _{\pm },\eta _{\pm })+z_0\int_{z_0}^B\gamma _1^z(\overline{\zeta
_{\pm }\xi _{\pm }})\widetilde{\gamma }_4^l(\zeta _{\pm },\eta _{\pm }) \\
C_2(\xi _{\pm };\eta _{\pm }) &=&-c_{{\bf Q}}\int_{c_{{\bf Q}}}\left[ \gamma
_1^c(\overline{\zeta _{\pm }\eta _{\pm }})\gamma _1^h(\xi _{\pm },\zeta
_{\pm })+\gamma _2^c(\overline{\zeta _{\pm }\eta _{\pm }})\gamma _2^h(\xi
_{\pm },\zeta _{\pm })\right] \\
Z_2(\xi _{\pm };\eta _{\pm }) &=&z_0\int_{z_0}^A\left[ \gamma _4^{\widetilde{%
z}}(\overline{\zeta _{\pm }\xi _{\pm }})\gamma _2^l(\zeta _{\pm },\eta _{\pm
})+\gamma _4^z(\overline{\zeta _{\pm }\xi _{\pm }})\widetilde{\gamma }%
_1^l(\zeta _{\pm },\eta _{\pm })-2\gamma _4^z(\overline{\zeta _{\pm }\xi
_{\pm }})\gamma _2^l(\zeta _{\pm },\eta _{\pm })\right] \\
&&\ +z_0\int_{z_0}^B\left[ \gamma _1^{\widetilde{z}}(\overline{\zeta _{\pm
}\xi _{\pm }})\gamma _4^l(\zeta _{\pm },\eta _{\pm })+\gamma _2^z(\overline{%
\zeta _{\pm }\xi _{\pm }})\widetilde{\gamma }_4^l(\zeta _{\pm },\eta _{\pm
})-2\gamma _2^z(\overline{\zeta _{\pm }\xi _{\pm }})\gamma _4^l(\zeta _{\pm
},\eta _{\pm })\right] \\
\widetilde{Z}_2(\xi _{\pm };\eta _{\pm }) &=&z_{{\bf Q}}\int_{z_{{\bf Q}%
}}\left[ \gamma _2^{\widetilde{z}}(\overline{\zeta _{\pm }\xi _{\pm }})%
\widetilde{\gamma }_2^l(\zeta _{\pm },\eta _{\pm })+\gamma _3^{\widetilde{z}%
}(\overline{\zeta _{\pm }\xi _{\pm }})\widetilde{\gamma }_3^l(\zeta _{\pm
},\eta _{\pm })\right]
\end{eqnarray*}

\begin{eqnarray}
C_3(\xi _{\pm };\eta _{\pm }) &=&-c_0\int_{c_0}^A\gamma _4^c(\overline{\zeta
_{\pm },\eta _{\pm }})\gamma _3^h(\xi _{\pm };\zeta _{\pm
})-c_0\int_{c_0}^B\gamma _3^c(\overline{\zeta _{\pm },\eta _{\pm }})\gamma
_4^h(\xi _{\pm };\zeta _{\pm })  \nonumber \\
Z_3(\xi _{\pm };\eta _{\pm }) &=&z_{{\bf Q}}\int_{z_{{\bf Q}}}\left[ \gamma
_3^z(\overline{\zeta _{\pm }\xi _{\pm }})(\widetilde{\gamma }_2^l(\zeta
_{\pm },\eta _{\pm })-\gamma _1^z(\zeta _{\pm },\eta _{\pm }))+(\gamma _2^{%
\widetilde{z}}(\overline{\zeta _{\pm }\xi _{\pm }})-\gamma _1^{\widetilde{z}%
}(\overline{\zeta _{\pm }\xi _{\pm }}))\gamma _3^l(\zeta _{\pm },\eta _{\pm
})\right.  \nonumber \\
&&\ \ \ +\left. \gamma _1^z(\overline{\zeta _{\pm }\xi _{\pm }})(\widetilde{%
\gamma }_3^l(\zeta _{\pm },\eta _{\pm })-\gamma _3^l(\zeta _{\pm },\eta
_{\pm }))+(\gamma _3^{\widetilde{z}}(\overline{\zeta _{\pm }\xi _{\pm }}%
)-\gamma _3^z(\overline{\zeta _{\pm }\xi _{\pm }}))\gamma _1^l(\zeta _{\pm
},\eta _{\pm })\right]  \nonumber \\
\widetilde{Z}_3(\xi _{\pm };\eta _{\pm }) &=&z_{{\bf Q}}\int_{z_{{\bf Q}%
}}\left[ \gamma _3^{\widetilde{z}}(\overline{\zeta _{\pm }\xi _{\pm }})%
\widetilde{\gamma }_2^l(\zeta _{\pm },\eta _{\pm })+\gamma _2^{\widetilde{z}%
}(\overline{\zeta _{\pm }\xi _{\pm }})\widetilde{\gamma }_3^l(\zeta _{\pm
},\eta _{\pm })\right]  \nonumber \\
C_4(\xi _{\pm };\eta _{\pm }) &=&-c_0\int_{c_0}^A\gamma _3^c(\overline{\zeta
_{\pm },\eta _{\pm }})\gamma _3^h(\xi _{\pm };\zeta _{\pm
})-c_0\int_{c_0}^B\gamma _4^c(\overline{\zeta _{\pm },\eta _{\pm }})\gamma
_4^h(\xi _{\pm };\zeta _{\pm })  \nonumber \\
Z_4(\xi _{\pm };\eta _{\pm }) &=&z_0\int_{z_0}^A\left[ \gamma _4^z(\overline{%
\zeta _{\pm }\xi _{\pm }})\widetilde{\gamma }_4^l(\zeta _{\pm },\eta _{\pm
})+\gamma _4^{\widetilde{z}}(\overline{\zeta _{\pm }\xi _{\pm }})\gamma
_4^l(\zeta _{\pm },\eta _{\pm })-2\gamma _4^z(\overline{\zeta _{\pm }\xi
_{\pm }})\gamma _4^l(\zeta _{\pm },\eta _{\pm })\right]  \nonumber \\
&&\ \ \ +z_0\int_{z_0}^B\left[ \gamma _2^z(\overline{\zeta _{\pm }\xi _{\pm }%
})\widetilde{\gamma }_1^l(\zeta _{\pm },\eta _{\pm })+\gamma _1^{\widetilde{z%
}}(\overline{\zeta _{\pm }\xi _{\pm }})\gamma _2^l(\zeta _{\pm },\eta _{\pm
})-2\gamma _2^z(\overline{\zeta _{\pm }\xi _{\pm }})\gamma _2^l(\zeta _{\pm
},\eta _{\pm })\right]  \nonumber \\
\widetilde{Z}_4(\xi _{\pm };\eta _{\pm }) &=&z_0\int_{z_0}^A\widetilde{%
\gamma }_4^z(\overline{\zeta _{\pm }\xi _{\pm }})\widetilde{\gamma }%
_4^l(\zeta _{\pm },\eta _{\pm })+z_0\int_{z_0}^B\widetilde{\gamma }_1^z(%
\overline{\zeta _{\pm }\xi _{\pm }})\widetilde{\gamma }_1^l(\zeta _{\pm
},\eta _{\pm })
\end{eqnarray}
The vertices are now given by 
\begin{eqnarray}
\gamma _i^c(\eta _{\pm }) &=&g_0+Z_i(\eta _{\pm },\eta _{\pm })+\widetilde{Z}%
_i(\eta _{\pm },\eta _{\pm })  \nonumber \\
\gamma _i^z(\xi _{\pm }) &=&g_0+C_i(\xi _{\pm },\xi _{\pm })+\widetilde{Z}%
_i(\xi _{\pm },\xi _{\pm })  \nonumber \\
\gamma _i^{\widetilde{z}}(\xi _{\pm }) &=&g_0+C_i(\xi _{\pm },\xi _{\pm
})+Z_i(\xi _{\pm },\xi _{\pm })  \nonumber \\
\gamma _i^h(\xi _{\pm },\eta _{\pm }) &=&\gamma _i^c(\eta _{\pm })+C_i(\xi
_{\pm },\eta _{\pm })  \nonumber \\
\gamma _i^l(\xi _{\pm },\eta _{\pm }) &=&\gamma _i^z(\xi _{\pm })+Z_i(\xi
_{\pm },\eta _{\pm })  \nonumber \\
\gamma _i^l(\xi _{\pm },\eta _{\pm }) &=&\gamma _i^{\widetilde{z}}(\xi _{\pm
})+\widetilde{Z}_i(\xi _{\pm },\eta _{\pm })
\end{eqnarray}
and the regions of integration are defined by 
\begin{eqnarray}
\int_{c_0}^Af(\zeta _{+},\zeta _{-}) &=&\int\limits_{-|\xi _{+}|}^{|\xi
_{+}|}\int\limits_{-|\xi _{-}|}^{|\xi _{-}|}d\zeta _{+}d\zeta _{-}f(\zeta
_{+},\zeta _{-})  \nonumber \\
\int_{c_0}^Bf(\zeta _{+},\zeta _{-}) &=&\int\limits_{-|\widetilde{\xi }%
_{+}|}^{|\widetilde{\xi }_{+}|}\int\limits_{-|\widetilde{\xi }_{-}|}^{|%
\widetilde{\xi }_{-}|}d\zeta _{+}d\zeta _{-}f(\zeta _{+},\zeta _{-}) 
\nonumber \\
\int_{c_{{\bf Q}}}f(\zeta _{+},\zeta _{-}) &=&\int\limits_{-|\xi
_{+}|}^{|\xi _{+}|}\left( \int\limits_{\min \{0,\zeta _{-}^{(1)}\}}^{\max
\{0,\zeta _{-}^{(1)}\}}+\int\limits_{\min \{-|\xi _{-}|\text{sign}\zeta
_{+},\zeta _{-}^{(2)}\}}^{\max \{-|\xi _{-}|\text{sign}\zeta _{+},\zeta
_{-}^{(2)}\}}\right) \frac{d\zeta _{+}d\zeta _{-}}{\cos 2\varphi }\left| 
\frac{k_{+}k_{-}}{k_{+}^2+k_{-}^2+2\cos 2\varphi k_{+}k_{-}}\right|
\,f(\zeta _{+},\zeta _{-})  \nonumber \\
&&\ +\int\limits_{\min \{0,|\xi _{+}|\text{sign}\xi _{-}\}}^{\max \{0,|\xi
_{+}|\text{sign}\xi _{-}\}}d\zeta _{+}\left| \frac{\min (\cos 2\varphi
k_{+},p_{-})+p_{-}}{\cos 2\varphi k_{+}+\sin 2\varphi p_{-}}\right| f(\zeta
_{+},\xi _{-})  \nonumber \\
&&\ +\int\limits_{\min \{0,|\xi _{-}|\text{sign}\xi _{+}\}}^{\max \{0,|\xi
_{-}|\text{sign}\xi _{+}\}}d\zeta _{-}\left| \frac{\min (\cos 2\varphi
k_{-},p_{+})+p_{+}}{\cos 2\varphi k_{-}+\sin 2\varphi p_{+}}\right| f(\xi
_{+},\zeta _{-})  \nonumber \\
\int_{z_0}^Af(\zeta _{+},\zeta _{-}) &=&\int\limits_{-|\eta _{+}|}^{|\eta
_{+}|}d\zeta _{+}f(\zeta _{+},\eta _{-})+\int\limits_{-|\eta _{-}|}^{|\eta
_{-}|}d\zeta _{+}f(\eta _{+},\zeta _{-})  \nonumber \\
\int_{z_0}^Bf(\zeta _{+},\zeta _{-}) &=&\int\limits_{-|\widetilde{\eta }%
_{+}|}^{|\widetilde{\eta }_{+}|}d\zeta _{+}f(\zeta _{+},\widetilde{\eta }%
_{-})+\int\limits_{-|\widetilde{\eta }_{-}|}^{|\widetilde{\eta }_{-}|}d\zeta
_{+}f(\widetilde{\eta }_{+},\zeta _{-})  \nonumber \\
\int_{z_{{\bf Q}}}f(\zeta _{+},\zeta _{-}) &=&\int\limits_{-|\eta
_{+}|}^{|\eta _{+}|}\left( \int\limits_{\min \{0,|\eta _{-}|\text{sign}\zeta
_{+}\}}^{\max \{0,|\eta _{-}|\text{sign}\zeta _{+}\}}+\int\limits_{\min
\{\zeta _{-}^{(1)},\zeta _{-}^{(2)}\}}^{\max \{\zeta _{-}^{(1)},\zeta
_{-}^{(2)}\}}\right) d\zeta _{+}d\zeta _{-}\left| \frac{k_{+}k_{-}}{%
k_{+}^2+k_{-}^2+2\cos 2\varphi k_{+}k_{-}}\right| f(\zeta _{+},\zeta _{-})
\end{eqnarray}
where 
\begin{eqnarray*}
k_{\pm } &=&\Lambda \text{sign}(\zeta _{\pm })\exp (-|\zeta _{\pm }|) \\
L(k_{\pm }) &=&\text{sign}(k_{\pm })\ln |\Lambda /k_{\pm }| \\
\zeta _{-}^{(1)} &=&-L[k_{+}/\cos 2\varphi ] \\
\zeta _{-}^{(2)} &=&-L[k_{+}/\cos 2\varphi ]
\end{eqnarray*}

{\sc Figure captions}

Fig.1. Diagrams (bubbles) for noninteracting susceptibilities near $q=0$ and 
${\bf q}={\bf Q}${\em \ }in (a) Peierls channel (b) Cooper channel. Solid
and dashed lines correspond to the electron Green functions near A and B
singularities with the spectra (\ref{eka}) and (\ref{ekb}) respectively.

Fig.2. The momentum dependence for (a) noninteracting susceptibility and (b)
noninteracting Cooper response at $t^{\prime }/t=0.3$

Fig.3. The vertices $\gamma _i$ ($i=1...4$). The solid lines inside the
circles show which incoming and outgoing particles have the same spin
projection.

Fig.4. The ferromagnetic (solid line), antiferromagnetic (long-dashed line),
and $d$-wave superconducting (short-dashed line) susceptibilities for the
two-patch model with (a) $t^{\prime }/t=0.15;$ $g_0=0.10$ (b) $t^{\prime
}/t=0.45;$ $g_0=0.10$ (c) $t^{\prime }/t=0.30;$ $g_0=0.01$

Fig.5 The phase diagram for the two-patch model in $\mu $-$t^{\prime }/t$
coordinates for $g_0=0.1$ ($U=3.95t$). Dotted line is the mean-field
boundary for antiferromagnetic phase, dot-dashed line for ferromagnetic one.

Fig.6. The phase diagram for the two-patch model in $g_0$-$t^{\prime }/t$
coordinates at Van Hove filling ($\mu =0$).

Fig.7. The representation of the vertex in the parquet approach as a sum of
bricks for the Cooper ($C$) and zero-sound ($Z,\widetilde{Z}$) channels.

Fig.8. The phase diagram from parquet equations in $\mu $-$t^{\prime }/t$
coordinates for $g_0=0.1$ ($U=3.95t$). The lines are the same as in Fig.5.

Fig.9. The phase diagram from parquet equations for Van Hove filling ($\mu
=0 $) in $g_0$-$t^{\prime }/t$ coordinates.

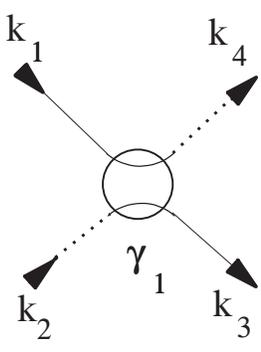
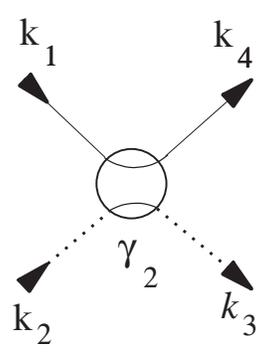
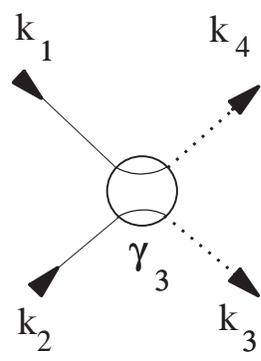
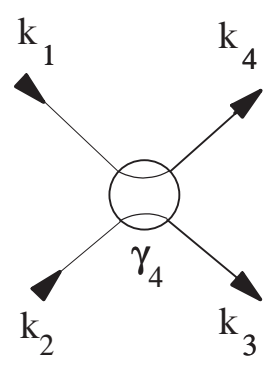
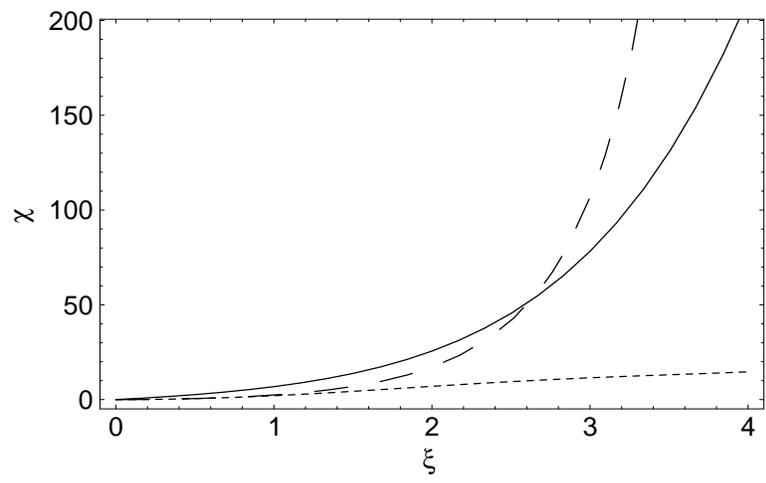
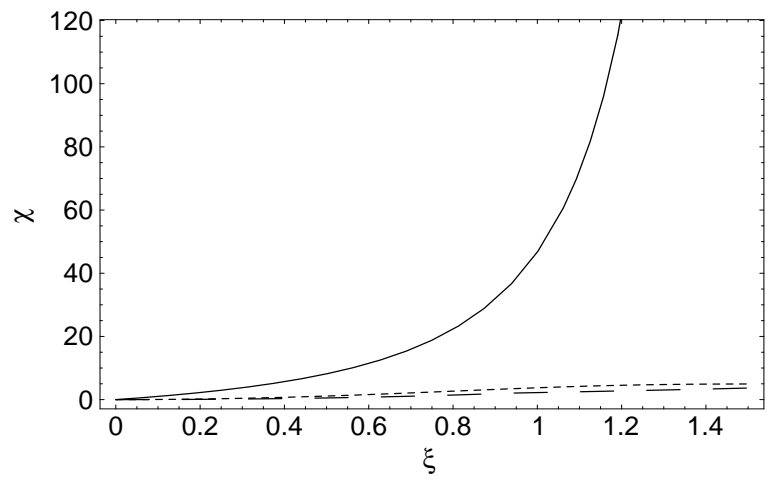
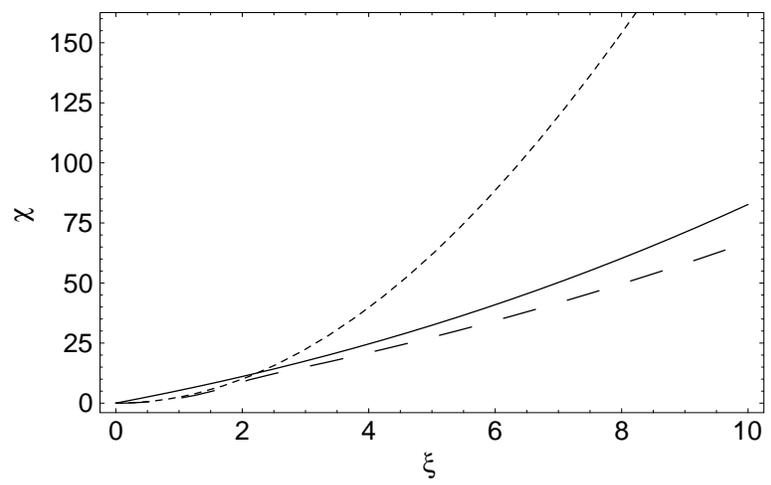
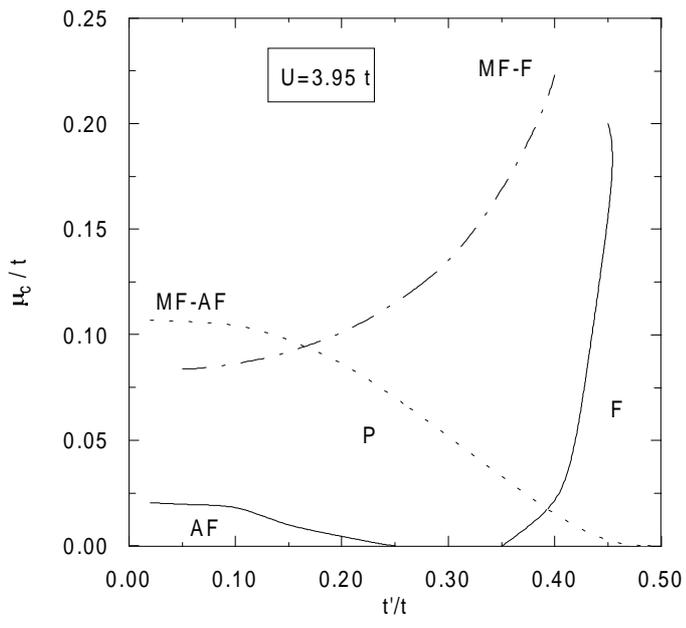
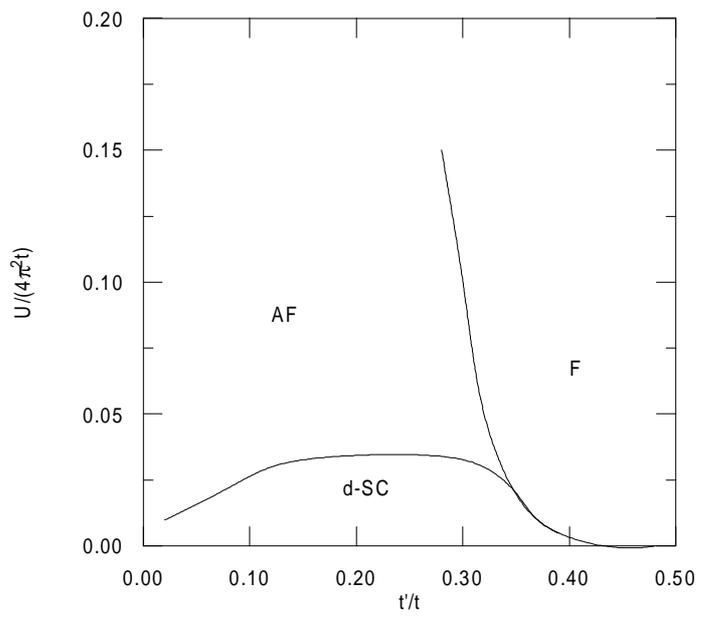
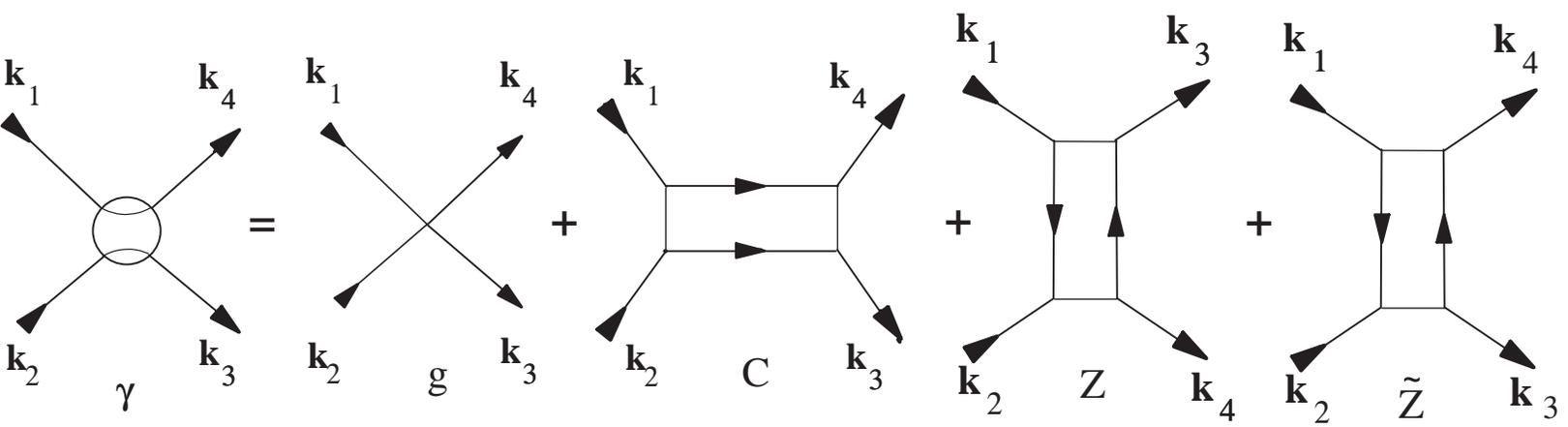
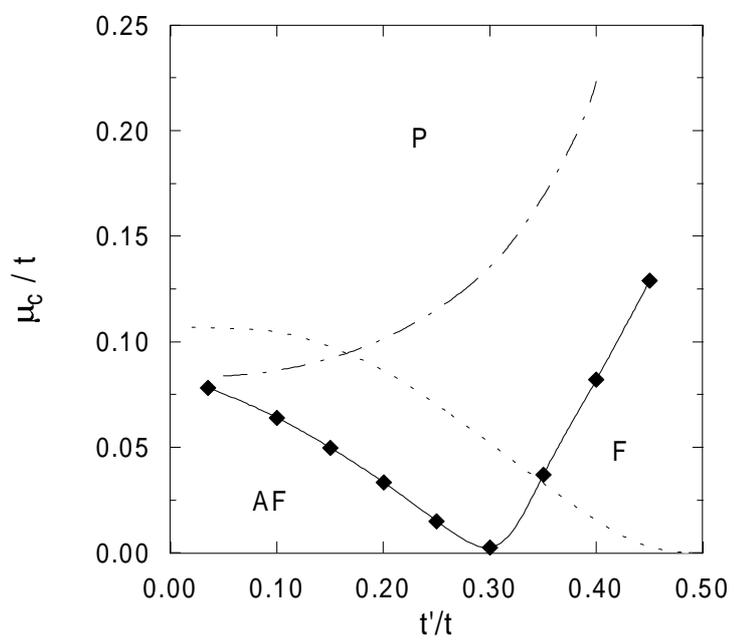
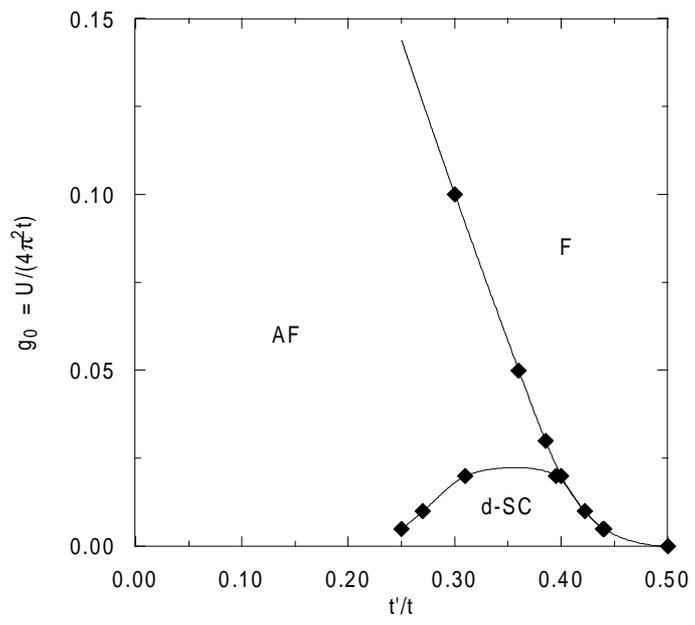
\end{document}